# Highly Efficient Ion Rejection by Graphene Oxide Membranes via Ion-controlling Interlayer Spacing

Liang Chen[1,2,4†], Guosheng Shi[2†], Jie Shen[3†], Bingquan Peng[1†], Bowu Zhang[2], Yuzhu Wang[2], Fenggang Bian[2], Jiajun Wang[1], Deyuan Li[2], Zhe Qian[1], Gang Xu[1], Guoquan Zhou[4], Minghong Wu[1*], Wanqin Jin[3*], Jingye Li[2*] and Haiping Fang[2*]

[1]Shanghai Applied Radiation Institute, Shanghai University, Shanghai 200444, China.
[2]Division of Interfacial Water, Key Laboratory of Interfacial Physics and Technology and Shanghai Synchrotron Radiation Facility, Shanghai Institute of Applied Physics, Chinese Academy of Sciences, Shanghai 201800, China.
[3]State Key Laboratory of Materials-Oriented Chemical Engineering, Jiangsu National Synergetic Innovation Center for Advanced Materials, Nanjing Tech University, 5 Xinmofan Road, Nanjing 210009, China.
[4]Zhejiang Provincial Key Laboratory of Chemical Utilization of Forestry Biomass, Zhejiang A&F University, Lin'an, Zhejiang 311300, China.

[†] These authors contributed equally to this work
*Corresponding author. E-mail: fanghaiping@sinap.ac.cn (H.-P. F.); mhwu@mail.shu.edu.cn (M.-H. W.); wqjin@njtech.edu.cn (W.-Q. J.) or lijingye@sinap.ac.cn (J.-Y. L.)

**Because they may provide ultrathin, high-flux, and energy-efficient membranes for precise ionic and molecular sieving in aqueous solution[1-5], GO membranes (partially oxidized, stacked sheets of graphene)[6] have shown great potential in water desalination and purification[7-9], gas and ion separation[10-13], biosensors[14], proton conductors[15], lithium-based batteries[16] and super-capacitors[17]. Unlike carbon nanotube (CNT) membranes, in which the nanotube pores have fixed sizes[18-20], the pores of GO membranes – the interlayer spacing between GO sheets – are of variable size. This presents a challenge for using GO membranes for filtration. Despite the great efforts to tune and fix the interlayer spacing[21-25], it remains difficult both to reduce the interlayer spacing sufficiently to exclude small ions while keeping this separation constant against the tendency of GO membranes to swell when immersed in aqueous solution[25], which greatly affects the applications of GO membranes. Here, we demonstrate experimentally that highly efficient and selective ion rejection by GO membranes can be readily achieved by controlling the interlayer spacing of GO membranes using cations ($K^+$, $Na^+$, $Ca^{2+}$, $Li^+$ and $Mg^{2+}$) *themselves*. The interspacing can be controlled with precision as small as 1 Å, and GO membranes controlled by one kind of cation can exclude other cations with a larger hydrated volume, which can only be**

**accommodated with a larger interlayer spacing. First-principles calculations reveal that the strong noncovalent cation-π interactions between hydrated cations in solution and aromatic ring structures in GO are the cause of this unexpected behavior. These findings open up new avenues for using GO membranes for water desalination and purification, lithium-based batteries and super-capacitors, molecular sieves for separating ions or molecules, and many other applications.**

There have been efforts to tune the interlayer spacing. For example, it can be widened to increase the permeability by intercalating large nanomaterials[21,22] as well as by cross-linking large and rigid molecules[23]. Reducing GO membranes can lead to sharp decrease of interlayer spacing, but to be highly impermeable to all gases, liquids and aggressive chemicals[24,25]. It remains difficult both to reduce the interlayer spacing sufficiently to exclude small ions while keeping this separation constant against the tendency of GO membranes to swell when immersed in aqueous solution[25]. This limits the potential of such membranes for separating ions from bulk solution, or sieving ions of a specific size range from a mixed salt solution, such as the most common ions in seawater as well as those in the electrolytes of lithium-based batteries and super-capacitors: $Na^+$, $Mg^{2+}$, $Ca^{2+}$, $K^+$ and $Li^+$[1,25]. Herein, combining experimental observations and theoretical calculation, we show that it is cations (i.e., $K^+$, $Na^+$, $Ca^{2+}$, $Li^+$, and $Mg^{2+}$) themselves that can determine and fix the size of nanochannels as small as 1 Å precision in GO membranes.

Freestanding GO membranes, prepared from GO suspension by the drop-casting method (Details in PS1 of Supplementary Information (SI)), were immersed in KCl, NaCl, $CaCl_2$, LiCl and $MgCl_2$ solutions of 0.25 mol/L (M) concentration for one hour. Then the wet membranes, saturated with salt solution, were taken out and analyzed by X-ray diffraction (XRD). There were clear shifts of the interlayer spacing (indicated by Bragg peaks, Fig. S3A) relative to that for pure water. The spacing for immersion in pure water was 12.8 ±0.2 Å, which is consistent with early reports[1,23]. For ionic solutions the spacings were 11.4 ±0.1 Å, 12.1 ±0.2 Å, 12.9 ±0.2 Å, 13.5 ±0.2 Å and 13.6 ±0.1 Å for KCl, NaCl, $CaCl_2$, LiCl and $MgCl_2$ solution, respectively (Fig. 1C). Thus, the order of the spacing is $MgCl_2$ > LiCl > $CaCl_2$ > pure water > NaCl > KCl. We noted that the interlayer spacing of 11.4 Å in KCl solution was even smaller than the narrowest interlayer spacing (~13 Å) of GO membranes in aqueous solution reported in literature[1,23,26]. The interlayer spacings were further confirmed using the two-dimensional (2D) synchrotron wide-angle X-ray scattering (WAXS) patterns (Details in PS3 of SI), which shows the same order of ion-controlling interlayer spacing as our experimental results from the XRD detection.

We also analyzed the solution adsorption of freestanding GO membranes with different ions. The wet mass of the freestanding GO membranes for KCl, NaCl, $CaCl_2$,

LiCl and MgCl$_2$ solution after removal of the solution on the surface by centrifugation was 2.4, 3.6, 3.0, 3.6 and 3.1 times of the dry mass of pristine GO membranes, respectively (Fig. 2A). We can see that, for the ions with the same valence, the order of these salt solutions adsorption is consistent with that for the interlayer spacings controlled by the corresponding salt solutions. For the salt solution adsorption of the ions with different valences, we attribute the order to the existence of ripples in these membranes (Details in PS3 of SI). The dry mass after drying at 60 °C for 6 hours is, for NaCl, CaCl$_2$, LiCl and MgCl$_2$, larger than that of pristine dry membranes, indicating that the salt is still retained. For KCl, on the other hand, the dry mass is approximately equal to that of the pristine dry membrane (Fig. 2A). This indicates that very little KCl penetrated into the GO membranes, which has been further confirmed by X-ray photoelectron spectroscopy (XPS) (Fig. S5). Thus, the membrane immersed in KCl solution rejected most of K$^+$ itself but still allowed the water to penetrate the membrane.

Thus one kind of cation can "control" the interlayer spacing and exclude other cations if they require a larger interlayer spacing. Considering that the KCl solution makes the narrowest interlayer spacing, we firstly soaked the freestanding GO membranes in pure water and next immersed them in KCl solution. Then, we added the same concentration solutions of other ions. The XRD spectra for those mixed solutions (KCl + M, M = NaCl, CaCl$_2$, LiCl, or MgCl$_2$) show only very small differences from the spectrum with pure KCl solution (fig. S3B). The corresponding interlayer spacings for KCl + M, with M = NaCl, CaCl$_2$, LiCl, or MgCl$_2$ are 11.4 ±0.2 Å, 11.4 ±0.1 Å, 11.2 ±0.2 Å and 11.2 ±0.1 Å, respectively (Fig. 1D). These results clearly show that K$^+$ can effectively fix the interlayer spacing at about 11 Å, resulting in rejection of other cations (including K$^+$ itself) in mixed solutions.

Furthermore, both the wet and dry masses of GO membranes immersed in mixed solution are all very close to that in pure KCl solution (Fig. 2B). This also suggests that the ions cannot permeate into the membranes after they have been "controlled" by the KCl solution – but water still can. This means that GO membranes treated with KCl solution may be useful for desalination technologies. We have also tested "control" of interlayer spacing by NaCl/CaCl$_2$ and found that the Na$^+$ and Ca$^{2+}$ also fixed the interlayer spacings and exclude the other cations if they require a larger interlayer spacing (Details in PS5 of SI).

In order to further demonstrate the controlling effects of K$^+$, GO membranes supported by ceramic substrates were fabricated and used for ion permeation tests. GO layers were deposited uniformly on Al$_2$O$_3$ substrate (Fig. 3A); SEM imaging (Fig. 3C) shows that these layers are defect-free, with a submicrometer thickness of about 300 nm, which is critical for a highly efficient separation process[5,27,28]. The ion permeation tests were carried out as shown in PS6 of SI. For GO membranes without KCl treatment, Na$^+$ could be transported at a permeation rate of 0.72 mol m$^{-2}$ h$^{-1}$ (Table in Fig. 3). However, after being treated with KCl solution, the permeation rate of $Na^+$ was reduced by a factor of more than 360. The water flux decreased owing to the fact that the GO interlayer spacing was reduced after treatment with $K^+$ (Fig. 1C and 1D), which was also verified by adsorption experiments (Fig. 2B).

We have investigated the underlying physical mechanism for this behavior. Generally the interaction of oxygen groups in oxidized graphene regions of GO sheets with ions are regarded as the main interaction between the GO sheets and ions. But in our classical molecular dynamics simulations of GO nanosheets immersed in salt solution, there was no accumulation of cations around the oxygen groups, and the interaction energy of oxygen groups with cations is no stronger than that of hydrogen bonding (Details in PS7 of SI). Therefore, it cannot be the interaction between oxygen groups on the GO membranes spacing and cations that controls the interlayer spacing.

What then is responsible? It has been shown previously that the cation-π interaction[29] between hydrated cations and aromatic rings is strong enough to cause blockage of water transport through carbon nanotubes, as well as causing ion accumulations on the graphite surface[20,30]. Using density functional theory (DFT), we find that the adsorption energy of hydrated $Li^+$ in between two unoxidized graphene regions of GO sheets reaches -26.3 kcal/mol, which is about four times larger than the strength of hydrogen bonds. The adsorption energies of all the other monovalent hydrated ions ($K^+$ and $Li^+$) are larger than this value (Fig. 4) and we can expect that the adsorption energies of the divalent hydrated cations ($Ca^{2+}$ and $Mg^{2+}$) are even larger[30]. The interlayer distances with hydrated $Li^+$, $Na^+$ and $K^+$ decrease in the order of $Li^+$-$(H_2O)_6$ > $Na^+$-$(H_2O)_6$ > $K^+$-$(H_2O)_6$ (Details in PS8 of SI), which is consistent with our experiment. In fact, the idea to use the ions *themselves* to determine and fix the interface spacing of GO membranes was first revealed by our theoretical computation. Note that the hydrated $Cl^-$-π interaction is only 1/10 of the hydrated $Na^+$-π interaction[30], and so $Cl^-$ does not play a role in the spacing control.

Why, though, does $K^+$ reject itself? The interaction between hydrated $K^+$ and graphene sheets is comparable to the hydration energy of $K^+$ (Fig. 4). This is different from the cases for other cations, in which the interaction between hydrated cations ($Li^+$-$(H_2O)_6$ and $Na^+$-$(H_2O)_6$) and graphene sheets are clearly smaller than to the corresponding hydrated energy of these cations. This suggests that the hydration structure of $K^+$ is unstable when the hydrated $K^+$ enters between the GO sheets, compared with other cations ($Li^+$ and $Na^+$). Fig. 4A show that the hydrated structure of $K^+$ inside in the GO sheets is distorted. $K^+$ is directly adsorbed on one side of the graphene sheet, resulting in a narrow interlayer spacing, and then other $K^+$ ions are rejected.

In summary, facile and precise control of the interlayer spacing in GO membranes, down to 1 Å precision, and corresponding ion rejection has been experimentally achieved by simply adding one kind of cation, based on our understanding of the strong noncovalent cation-π interactions between hydrated cations and aromatic ring. We note that our previous DFT computations shows that other cations ($Fe^{2+}$, $Co^{2+}$, $Cu^{2+}$, $Cd^{2+}$, $Cr^{2+}$, and $Pb^{2+}$) have a much stronger cation-π interaction[30] with the graphene sheet, which suggests that a wider range of interlayer spacing to that observed here can be obtained by other ions. Thus, our finding brings closer GO-based applications such as water desalination and gas purification, solvent dehydration, lithium based batteries and super-capacitors, and molecular sieving.

**Acknowledgments:** We thank Drs. Philip Ball, Lingxue Kong, Jian Liu, Zhengchi Hou, Guanhong Lei and Haijun Yang for their constructive suggestions. The supports from NNSFC (Nos. 41430644, 21490585, 21476107, 11290164 and 11404361) and Supercomputer Center of CAS and BL16B1 beamline at SSRF are acknowledged.

**Author contributions:** H.-P. F. had the idea to control the interlayer spacing by ions based on cation-π interactions. H.-P. F., M.-H. W., W.-Q. J., J.-Y. L. and G.-S. S. conceived and designed the experiments and simulations. L. C., G.-S. S., J. S., B.-Q. P., G. X., B.-W. Z., Y.-Z. W., F.-G. B., J.-J. W., D.-Y. L. and Z. Q. performed the experiments. G.-S. S. and L. C. performed the simulations. G.-S. S., L. C., H.-P. F., J.-Y. L., W.-Q. J., M.-H. W., G. X. and G.-Q. Z. analyzed the data, G.-S. S., L. C., H.-P. F., W.-Q. J. and M.-H. W. co-wrote the paper. All authors discussed the results and commented on the manuscript.

**Author Information:** Reprints and permissions information is available at www.nature.com/reprints. The authors declare no competing financial interests. Readers are welcome to comment on the online version of the paper. Correspondence and requests for materials should be addressed to H.-P. F. (fanghaiping@sinap.ac.cn); M.-H. W. (mhwu@mail.shu.edu.cn); W.-Q. J. (wqjin@njtech.edu.cn) or J.-Y. L. (lijingye@sinap.ac.cn).

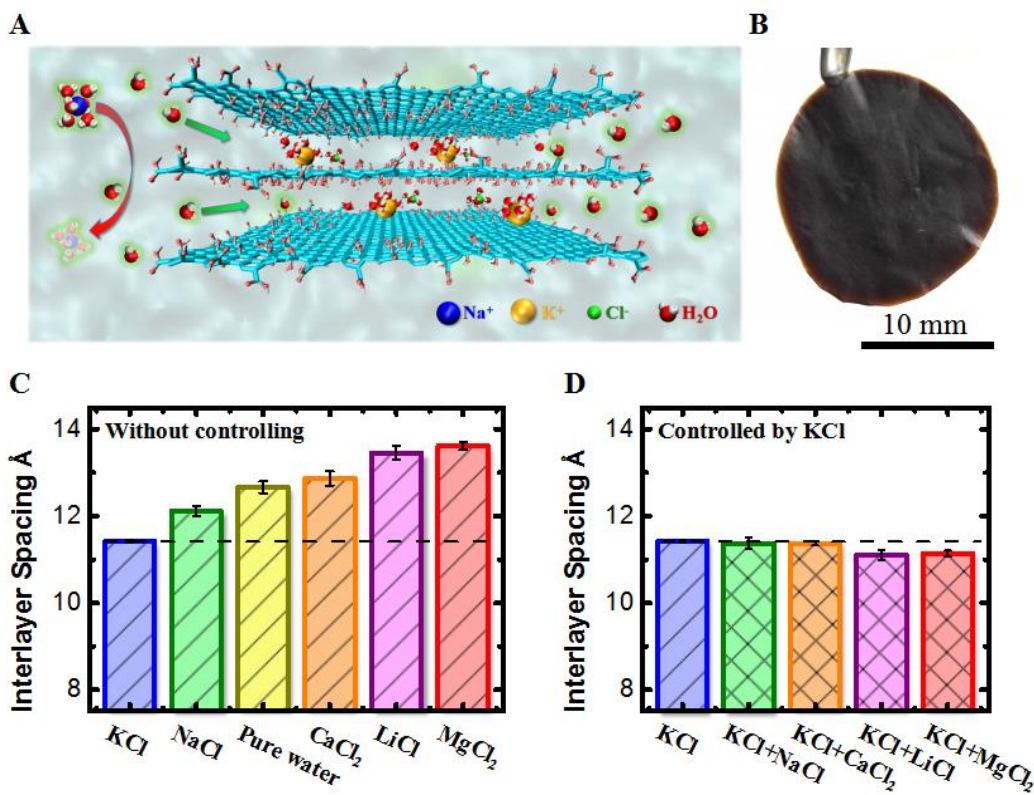

**Fig. 1. Interlayer spacings in freestanding GO membranes controlled by cations.** (**A**) A schematic that $K^+$ ions in the GO membranes fix the interlayer spacing to reject other cations while allowing pure water to penetrate. The left red curved arrow depicts that a hydrated $Na^+$ is rejected by the GO membrane with $K^+$ inside. (**B**) Photograph of a freestanding GO membrane prepared by drop-casting of a 5 mg/mL GO suspension. (**C**) Interlayer spacings for GO membranes immersed in pure water or various 0.25 mol/L (M) salt solutions. (**D**) Interlayer spacings of GO membranes, first soaked in KCl solution and then immersed in various salt solutions.

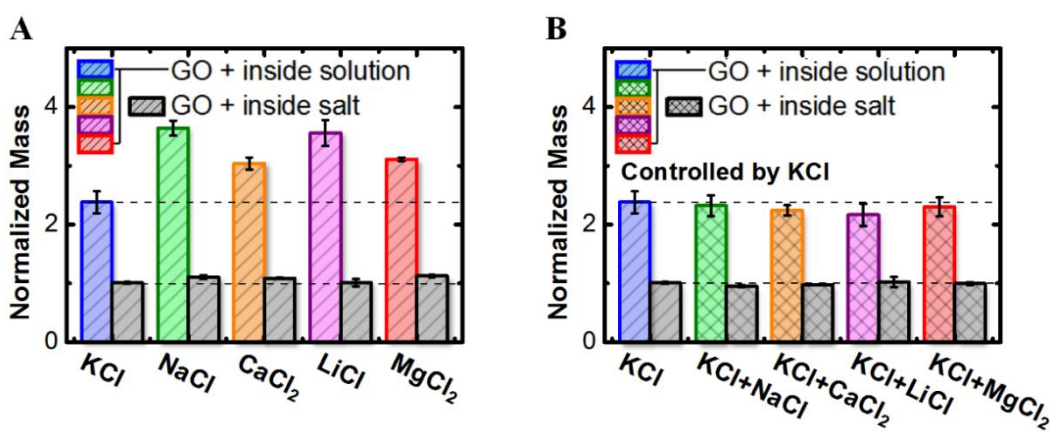

**Fig. 2. Solution adsorption of freestanding GO membranes with different ions.** (**A**) The wet and dry masses of the freestanding GO membranes after being immersed in various salt solutions, normalized by the corresponding dry mass of pristine GO membrane. The wet and dry masses are achieved after removal of solution on the surface by centrifugation and further drying at 60 °C for 6 hours, respectively. The two dashed lines correspond to the normalized dry mass of pristine GO membranes and the normalized wet mass for KCl. **(B)** The normalized wet and dry masses of GO membranes, first soaked in KCl solution, and then immersed in various salt solutions.

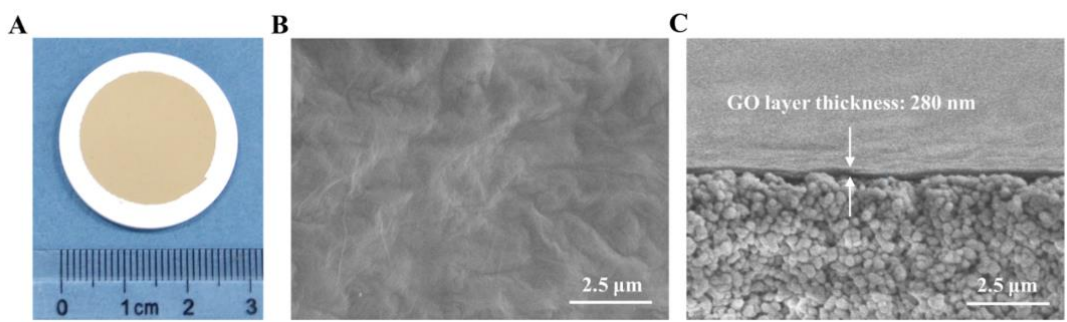

| Membrane | Concentration of Na$^+$ in feed solution after permeation (10$^{-3}$ mol/ L) | Na$^+$ permeation rate (mol m$^{-2}$ h$^{-1}$) | Water flux (L m$^{-2}$ h$^{-1}$) |
|---|---|---|---|
| GO without treatment | 11.69 | 0.72 | 0.85 |
| GO treated with KCl | <0.03* | <0.002 | 0.28 |

\*: below the detection limit

**Fig. 3. Ion permeation tests of GO membranes supported by ceramic substrates.** **(A)** Digital photo, **(B)** surface SEM image and **(C)** cross-section SEM image for the Al$_2$O$_3$ supported GO membrane. Table shows the results for Na$^+$ permeation tests.

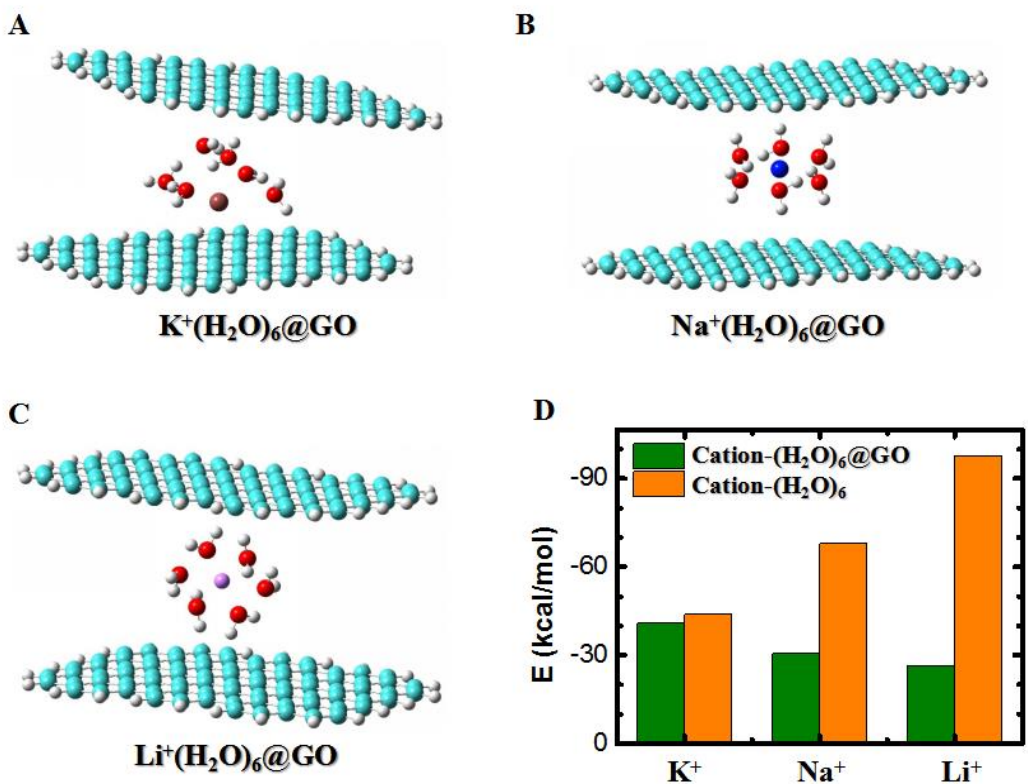

**Fig. 4. Density functional theory computation of the hydrated cations in between two unoxidized graphene regions of GO sheets.** The most stable optimized geometries of cation-$(H_2O)_6$@GO clusters, where the cation = $K^+$ **(A)**, $Na^+$ **(B)** and $Li^+$ **(C)**. **(D)** Hydration energies of cations with six water molecules (cation-$(H_2O)_6$, orange pillars) and adsorption energies of cation-$(H_2O)_6$ in between two unoxidized graphene regions of GO sheets (cation-$(H_2O)_6$@GO, green pillars).

**Supplementary Information:**
PS1: Materials and methods:
    Fabrication of freestanding GO membranes
    Fabrication of GO membranes supported by ceramic substrates
    Characterization methods
    Experimental setup for XRD detection
    Experimental setup for adsorption of salt solutions
    Theoretical computational methods:
        Molecular dynamics (MD) simulations
        Density functional theory (DFT) calculations
PS2: X-ray diffraction for freestanding GO membranes in solution
PS3: Synchrotron wide-angle X-ray scattering (WAXS) 2D patterns of freestanding GO membranes in dry state and solutions of 0.25 M KCl, NaCl, pure water, $CaCl_2$, LiCl, $MgCl_2$
PS4: XPS spectra of freestanding GO membranes: Pristine GO membrane and GO membrane immersed in KCl solution

PS5: Salt solution adsorption of freestanding GO membranes controlled by NaCl/CaCl$_2$

PS6: Ion permeation tests of GO membranes supported by ceramic substrates

PS7: Classic MD simulations on the GO sheets immersed in salt (KCl, NaCl, LiCl, CaCl$_2$ and MgCl$_2$) aqueous solution

PS8: DFT calculations on the hydrated cations in between two unoxidized graphene regions of GO sheets